\documentclass[prc,aps,twocolumn]{revtex4}
\usepackage{graphicx} 
\begin{document}

\title{New Properties of High Momentum Distribution of Nucleons in Asymmetric Nuclei.}

\author{Misak~M.~Sargsian}

\affiliation{Department of Physics, Florida International University, Miami, FL 33199 USA}

\date{\today}

\begin{abstract} 
Based on the recent experimental observations  of the dominance of tensor interaction in the 
$\sim$~250-600~MeV/c  momentum range of nucleons in nuclei,  the existence of 
 {\em two new properties}  for  high-momentum distribution of nucleons in  asymmetric nuclei is suggested.
The {\em first} property is the approximate scaling relation between proton and neutron high-momentum 
distributions weighted by their relative fractions in the nucleus. The {\em second} property is 
the  inverse proportionality of the strength  of  the  high-momentum distribution of protons and neutrons to the same 
relative fractions.
Based on these two properties the high-momentum distribution function  for asymmetric nuclei  has been modeled 
and  demonstrated that it describes reasonably well the high-momentum characteristics  of light nuclei. 
However,  the most surprising  result  is  
obtained  for neutron rich nuclei with large $A$, for which  {\em a substantial  relative  abundance} of 
high-momentum protons  as compared to  neutrons is predicted.
For example,  the model predicts  that in   Au  the relative fraction of protons with  momenta above  
$k_{F}\sim 260$~MeV/c is  50\% more   than  that of neutrons.  Such a situation may  have many  implications 
for different observations  in nuclear physics  related to the 
properties of a proton in neutron rich nuclei.
 \end{abstract}
\maketitle

\section{Introduction}
One of the exciting recent results in the studies of short-range properties of nuclei is the observation of the 
strong  (by factor of 20) dominance  of the  $pn$ short range correlations~(SRCs) in nuclei,   as compared to  $pp$ and $nn$ 
correlations, for internal momenta  of $\sim 250-600$~MeV/c\cite{isosrc,EIPsc}. 
This observation is understood\cite{isosrc, t2,Sch} 
based on the dominance of the tensor forces in the NN interaction at this momentum range corresponding to 
average nucleon separations of $\sim 1.1$~Fm.   The tensor interaction projects the NN SRC part of the 
wave function into the  isosinglet - relative angular momentum, $L=2$  state,  almost identical to the $D$-wave component of 
the deuteron wave function.  
At the same time,  $pp$ and $nn$ components of the NN SRC will be strongly suppressed since they 
are dominated by the central NN potential with relative $L=0$. 

In this work we explore 
the implication of the above observation on the properties of high momentum distribution of  nucleons in asymmetric nuclei. Two 
new features are predicted: {\em first}, that high momentum distributions of the proton and neutron weighted by their relative fractions  
are approximately equal (Sec~III) ; and {\em second}, for moderately asymmetric nuclei the high momentum distribution of  
nucleon is inverse proportional to 
its fraction in the nucleus (Sec~IV). In Sec.~V we demonstrate that these properties predict strikingly different high momentum tails for 
proton and neutron in neutron reach nuclei such as Gold.  
Section~VI discusses the results of  realistic calculations for light nuclei (up to $^{11}B$) which are in reasonable 
agreement with  the predicted properties  of high momentum distribution.  Furthermore, in  
Sec.~VII,  we discuss  the possibilities of  verification of the same properties  for heavy, neutron rich, nuclei through the probing 
the high momentum distribution of nucleons in semi-inclusive electro-nuclear reactions.  Section VIII discusses the restrictions of the model and 
the accuracy of the predictions.  We then discuss (Sec.~IX)  the possible implications of the new properties in different nuclear 
phenomena such as isospin dependence of the medium modification effects and properties of the proton in high density nuclear matter. 
This section also addresses the question of universality of the predicted features for any asymmetric  two-component Fermi system controlled only  
by short range interaction between the components.  The conclusions are given in Sec.~X.

\section{High Momentum Distribution of Nucleons in Nuclei and 2N SRCs} 
Due to the short-range nature of strong interactions, the property of an  A-nucleon 
bound state wave function, in which one of the nucleons has momentum $p$, such that  ${p^2\over 2m_N}\gg |E_{B}|$ 
(binding energy), is 
defined mainly by the 2N  interaction potential~($V_{NN}$)  at relative momenta $k\sim p$, i.e.: 
 $\Psi_{A}(p,p_2,p_3,\cdots p_A) \sim {V_{NN}(k)\over k^2} f(p_3,\cdots p_A)$, where 
 $\vec p_2\approx - \vec p \approx -\vec k$ 
 and $f(\cdots)$ is a smooth function of the momenta of non-correlated nucleons\cite{FS81,srcrev,Ciofi_Simula}. This 
 result follows from a dimensional analysis of the Lipmann-Schwinger type equations for A-nucleon system described by 
 $NN$  potential which decreases at large $k$ as  $V(k)\sim {1\over k^n}$, with $n>1$\cite{FS81,srcrev}. This asymptotic 
 form of the wave function leads to the  approximate relation for nucleon momentum distribution  at $p> k_{F}$, with $k_F$ being the characteristic  Fermi momentum of the nucleus: 
 \begin{equation}
 n^A({p}) \sim a_{NN}(A) \cdot n_{NN}({p})
\label{2NSRCmodel}
 \end{equation}
 where the full momentum distribution is normalized as $\int n^A({p})d^3p = 1$. 
 The parameter  $a_{NN}(A)$ can be interpreted as a probability 
 of finding NN SRC in the given nucleus $A$. The function, $n_{NN}({p})$ is the momentum distribution 
 in the NN SRC\cite{FS81,srcrev,Arrington:2011xs,FS88,FSDS}, where NN represents the combination of all 
 possible isospin pairs.

If, following the   above discussed dominance of tensor interactions,
we neglect   the   contributions from  $pp$ and $nn$  SRCs, one 
expects that in the range of $\sim k_{F}-600$~MeV/c the momentum distribution in the NN SRC is defined 
by $pn$ correlations only.  Using this and the local nature of SRCs one predicts:
\begin{equation}
  n_{NN}({p})\approx n_{pn}({p}) \approx  n_{d}({p}),
 \label{2=d}
 \end{equation}
where $n_d({p})$ is the deuteron  momentum distribution. 

For  further discussion we introduce the individual momentum distributions of 
proton ($n^{A}_{p}({p})$)  and neutron($n^A_n({p})$) such that:
\begin{equation}
n^A({p})  = {Z\over A} n^A_{p}({p}) +   {A-Z\over A}n^A_n({p})
\label{sum}
\end{equation}
and $\int n^A_{p/n}({p})d^3p = 1$.  Here the two terms in the sum represent the 
probability density of finding in the nucleus a proton or neutron with momentum $p$.

\section{ Approximate Scaling Relation} 
Integrating Eq.(\ref{sum}) within the momentum range of NN SRCs one observes that 
the terms in the sum give the  total probabilities of finding a proton  and a  neutron in the 
NN SRC.  Since the SRCs within our approximation consist only of the $pn$-pairs, 
the total probabilities  of finding proton and neutron in the SRC are equal
This is the reflection of the fact 
that in our approximation no other possibilities exist for NN SRCs. 
Furthermore within the approximation in which  one neglects the center of mass motion of the $pn$ SRCs 
one can make a stronger statement on the equality of integrands of the above integrals, i.e. in 
 $\sim k_F-600$~MeV/c region:
\begin{equation}
x_p\cdot n^{A}_{p}({p}) \approx x_n\cdot n^A_n({p}),
\label{p=n}
\end{equation}
where  $x_p = {Z\over A}$, $x_n = {A-Z\over A}$. This represents the {\em first} property,
according to which the momentum distributions of proton and neutron weighted by their respective 
fractions are approximately equal. 

\section{Fractional Dependence of High Momentum Components}
Using the high momentum relations of  (\ref{2NSRCmodel}) and (\ref{2=d}) for $n^A({p})$ 
and  the relation (\ref{p=n}) in Eq.(\ref{sum}) one obtains that in 
$\sim k_F-600$~MeV/c momentum range
 \begin{equation}
 n^{A}_{p/n}({p}) \approx {1\over 2 x_{p/n}} a_2(A,y)\cdot n_d({p}),
 \label{highn}
 \end{equation}
 where $a_{NN}(A) \approx a_{pn}(A,y)\equiv a_2(A,y)$   and  
 the nuclear asymmetry parameter  is defined as  $y= |x_n-  x_p|$.
 
Within the approximation in which  only  $pn$ SRCs are  included 
the parameter $a_{2}(A,y)$  satisfies to two limiting conditions:
(i) $a_2(A,0)$ is defined only by the nuclear density; and (ii) $a_2(A,1) = 0$ due 
to the neglection of   $pp$ and $nn$  SRCs. This allows us to represent  $a_{2}(A,y)$ as:
\begin{equation}
a_{2}(A,y) = a_2(A,0)\left[1 - \sum\limits_{j=1}^n b_j \mid x_n-x_p\mid^j\right],
\label{a2s}
\end{equation}
with the condition $\sum\limits_{j=1}^n b_j = 1$ to satisfy the limiting condition (ii). The latter relation 
indicates 
that it is always  possible to satisfy an inequality: $\sum\limits_{j=1}^n b_j \mid x_n-x_p\mid^j \ll 1$ in
which case one can formulate  the  {\em second} property of the high momentum 
distribution:  that,  according to Eq.(\ref{highn}), the probability of proton or neutron being in high 
momentum NN correlation is inverse proportional to their  relative fractions ($x_p$ or $x_n$)  in the nucleus.

\section{Relative Number of High Momentum Protons and Neutrons}
The most important prediction that follows from the {\em second property}  is that the relative number 
of high momentum  protons and neutrons became  increasingly  {\em unbalanced}  with an  increase 
of the nuclear  asymmetry,  $y$. To quantify this prediction, using  Eq.(\ref{highn}) one can calculate the   fraction 
of the   nucleons having  momenta $\ge k_{F}$ as:
\begin{equation}
P_{p/n}(A,y) \approx  {1\over 2 x_{p/n}} a_2(A,y)\int\limits_{k_F}^{\infty} n_d(p) d^3p,
\label{fraction}
\end{equation}
where we extended the upper limit of integration to infinity assuming smaller overall contribution 
from the momentum range of $\ge 600$~MeV/c.
The results of  the calculation  of these fractions for medium to heavy nuclei, 
using the  estimates of $a_2(A,y)$ from  Ref.\cite{FSDS,Kim1,Kim2,Fomina2,proa2} 
and $k_{F}$ from Ref.\cite{Moniz} are given in Table.\ref{table1}. 
\begin{table}[t]
\caption{Fractions of high momentum protons and neutrons in nuclei A.} 
\centering
\begin{tabular}{llllll}
\hline\hline 
A & $P_p(\%)$  & $P_n(\%)$ & A & $P_p(\%)$  & $P_n(\%)$ \\  [0.5ex] 
 \hline
12  & 20 & 20 &
56  & 27 & 23 \\
27  & 24 & 22 &
197 & 31 & 20 \\
\hline
\end{tabular}
\label{table1}
\end{table}
As it  follows from the table with  the increase of  the asymmetry the imbalance between the high momentum 
fractions of proton and neutron grows. 
For example,   in the Gold,  the relative fraction of high momentum ($\ge k_{F}$)  protons  is  50\% more  
than that of the neutrons.

\section{High Momentum Features of Light Nuclei} One can check the validity of 
the above two (Eqs.(\ref{p=n}) and (\ref{highn})) observations for light nuclei for which  it is possible to 
perform realistic calculations based on the Faddeev equations for A=3 systems\cite{Bochum},
Correlated Gaussian Basis(CGB) approach\cite{CGB}  as well as  Variational Monte Carlo method(VMC)\cite{VMC}  
for light nuclei  A (recently being available for up to $A =12$ \cite{VMCpc2,VMCpc}). 

\begin{figure}[ht]
\centering\includegraphics[scale=0.45]{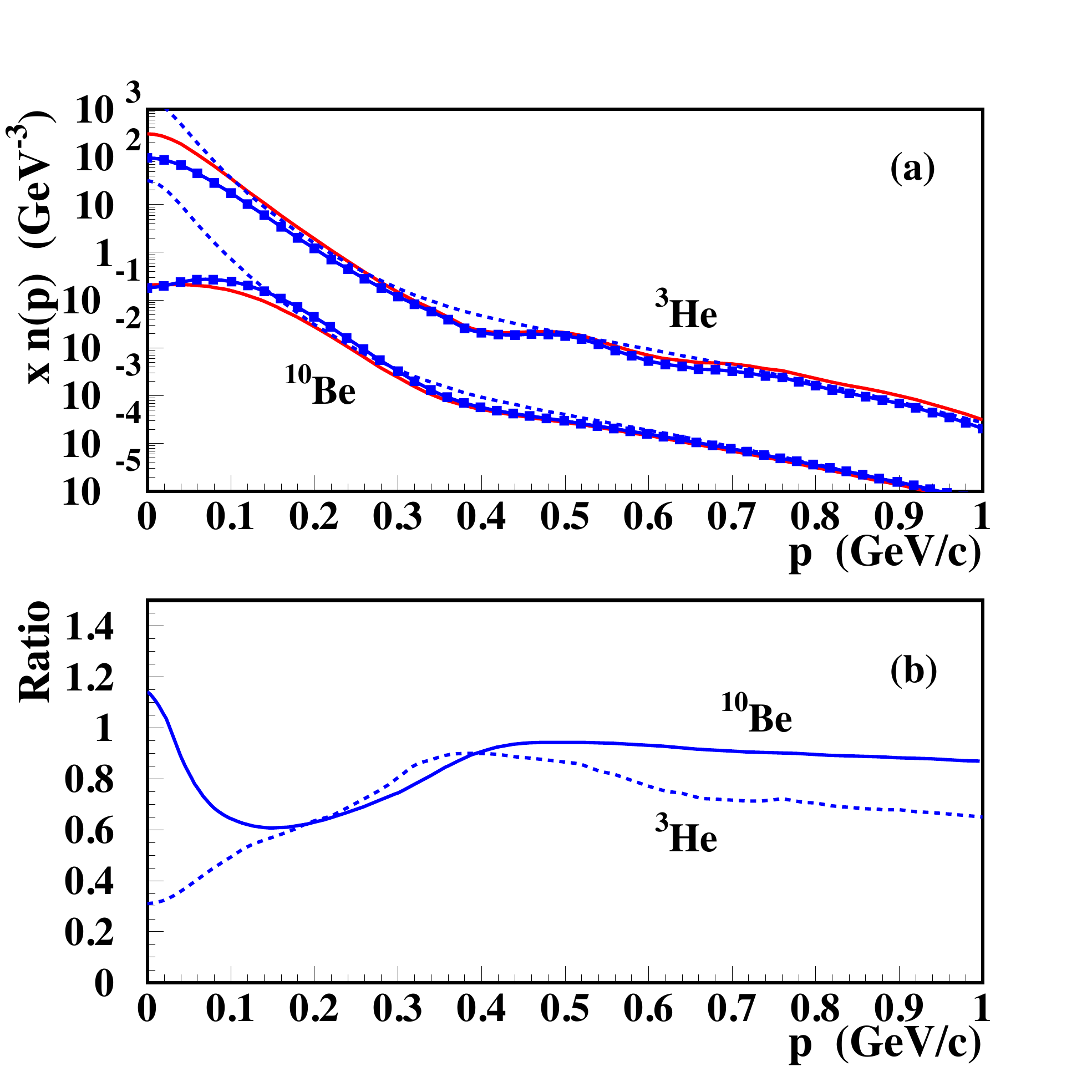}
\caption{(color online) (a) The momentum distributions of proton  and neutron weighted by $x_p$ and $x_n$ respectively.
The doted lines represent the prediction for the  momentum distribution according to Eq.(\ref{highn}).  \ 
 (b)  The $x_{p/n}$ weighted ratio 
of  neutron to  proton  momentum distributions.  See the text for details.}
\label{He3_Be10}
\end{figure}

First,  we check the validity of Eq.(\ref{p=n}) which is presented in Fig.\ref{He3_Be10} for 
$^3He$ nucleus,  based on the solution of  Faddeev equation\cite{Bochum},   and 
for $^{10}Be$  based on VMC calculations\cite{VMC}.  In both cases the 
Argonne V18\cite{V18} potential is used for NN interaction.
The solid  lines with and without squares in Fig.\ref{He3_Be10}(a)  represent 
neutron and proton momentum distributions for both nuclei weighted by their respective 
${x_n}$ and ${x_p}$ factors.

As one can see for $^3He$,  the proton momentum distribution 
dominates the neutron  momentum distribution at small momenta reflecting the fact that in the mean field 
the probability of finding proton is larger than neutron   just because there are twice as much protons in $^3He$. 
The same is true for $^{10}Be$ for which now the neutron momentum distribution dominates at small momenta.
However   at  $\sim 300$~MeV/c  for both nuclei, the proton and neutron momentum distributions become close  to each other 
up to the  internal momenta of $600$MeV/c. This is the region dominated by tensor interaction.

This effect is more visible for the ratios  of weighted  n-  to p- momentum 
distributions in Fig.\ref{He3_Be10}(b), demonstrating  that the approximation of Eq.(\ref{p=n}) in the range of 
$300-600$ MeV/c  is good  on the level of 15\%.     Note that the similar features  present for 
all other asymmetric nuclei calculated  within the VMC method in Ref.\cite{VMCpc2,VMCpc}.

Next, we check the the validity of Eq.(\ref{highn}). For this we use the estimates of $a_2$  for $^3He$ and 
$^{10}Be$ from  Refs.\cite{proa2,Fomina2} and  the deuteron momentum distribution $n_d$ calculated using 
the same Argonne V18 NN potential\cite{V18}.
The calculations based on Eq.(\ref{highn}) are given by doted lines in Fig.\ref{He3_Be10}(a). 
As it follows from  these comparisons, the model of Eq.(\ref{highn}) works rather well starting at 200~MeV/c up to the 
very large momenta $\sim 1$~GeV/c. This reflects the fact that the center of mass motion effects and  higher partial waves 
in 2N- as well as 3N- SRCs are not dominant in light nuclei

The final prediction which we check  is the one following from  Eq.(\ref{fraction}) according to which 
the smallest component should be more energetic in the asymmetric nuclei.  Namely, one expects more energetic 
neutron than proton in $^3He$  and  the opposite result for neutron rich nuclei.   This expectation is confirmed 
for $p-$ and $n-$ kinetic energies of all nuclei calculated within Faddeev equation, CGB approach and  VMC method
(see Table \ref{table2}).

\begin{table}[t]
\caption{Kinetic energies (in MeV) of proton and neutron\\} 
\centering 
\begin{tabular}{l l l l c }
\hline\hline 
A & y & $E_{kin}^p$  &   $E_{kin}^n$  &   $E_{kin}^p-E_{kin}^n$\\  [0.0ex] 
\hline 
$^{8}$He     \    & 0.50  \  & 30.13 \  &  18.60 \  & 11.53     \\ 
$^{6}$He     \    & 0.33  \  & 27.66 \  & 19.06  \  &  8.60  \\ 
$^{9}$Li       \    & 0.33  \  & 31.39 \  &  24.91 \  & 6.48    \\ 
$^{3}$He     \    & 0.33  \  & 14.71 \  & 19.35  \  &  -4.64   \\ 
$^{3}$He\cite{Bochum} \ & 0.33 \ & 13.70 \ & 18.40 \ &  -4.7 \\
$^{3}$He\cite{CGB} \ & 0.33 \ & 13.97 \ & 18.74 \ &  -4.8 \\
$^{3}$H       \    & 0.33  \  & 19.61   \  &  14.96 \  &   4.65   \\ 
$^{8}$Li       \    & 0.25  \  & 28.95 \  &  23.98 \  & 4.97   \\ 
$^{10}$Be   \    & 0.2    \   & 30.20 \  &  25.95 \  & 4.25    \\
$^{7}$Li       \    & 0.14  \  & 26.88 \  &  24.54 \  & 2.34   \\ 
$^{9}$Be     \    & 0.11  \   & 29.82 \  &  27.09 \  & 2.73    \\
$^{11}$B     \    & 0.09   \  \  \ & 33.40 \  \  \ &  31.75 \ \ \  &   1.65    \\ 
 \hline 
\end{tabular}
\label{table2} 
\end{table}

Thus we conclude that all the  observations concerning the  features of high momentum distribution 
in asymmetric nuclei are in reasonable agreement with the results following 
from the realistic   wave functions of light nuclei.

\section {High Momentum Features of Heavy Nuclei}
Presently, no ab-initio calculations exist for heavy nuclei for the predictions of Eqs.(\ref{p=n}) and (\ref{highn}) 
to be checked. 
However, these properties can be checked experimentally  in semi-inclusive 
nucleon knock-out $A(e,e'N)X$ reactions on asymmetric nuclei in which  the  momentum distribution of 
the nucleon can be probed if final state interactions(FSI) are  in control. Such a control can be achieved 
at large $Q^2>1$~GeV$^2$ kinematics  in  which case it was demonstrated 
that the FSI effects can be  estimated reasonably well within 
eikonal approximation (see e.g. \cite{Garrow:di} and references therein).   The first such experimental  verification 
for heavy nuclei  is currently   underway in quasi-elastic $A(e,e,p)X$ measurement at Jefferson Lab, where  
the ratio of  high momentum fractions of  nucleons in $^{56}Fe$ and $^{208}Pb$  to that of $^{12}C$ is extracted.  The
results\cite{EIPpc} are in reasonably good agreement with the prediction of  Eq.(\ref{fraction}) (Table 1)  and 
they are currently being prepared for publication.  

It is worth noting  that there is a possibility of designing a host of  new  $(e,e^\prime N)$ experiments with 
asymmetric nuclei  at specific kinematics in which  $x_{Bjorken}> 1$ and  
$|p_{m}^z| - {q_0\over q_v}(E_m+{p_m^2\over 2 M_{A-1}}) > k_F$,  where $p_m$, $E_m$,  $q_0$ and $q_v$  
are missing momentum, missing energy,  transferred  energy and  transferred momentum in the reaction 
(see e.g. Ref.\cite{gea,ms01} for details) in which case it is possible to extract the high momentum distribution of 
nucleons with minimal distortion due to FSI effects. Such measurements will allow to check also the 
predictions of Eqs.(\ref{p=n}) and (\ref{highn}). Moreover the $(e,e^\prime N)$ experiments will allow to extract 
nuclear spectral functions which contain additional information on the structure of SRCs, such as correlation 
between missing energy and missing momentum. One of the first measurements\cite{c12spectral} of 
the nuclear spectral function at SRC  region confirmed the high potential of the   $(e,e^\prime N)$ reactions in correlation studies.

\section{Restriction of the Model:}       
In the  above  made observations we neglected the $pp$ and $nn$ SRCs which are present in non-tensor (e.g. S=0-state) 
as well as the $T=1$, $S=1$  part of the NN interactions.  These contributions are expected to increase with A.  We
also  neglected the C.M. motion of  $pn$ SRCs.  It is rather well established that, for $A\ge 12$, 
in the momentum range of $k_F < p < 600$~MeV/c the C.M. momentum of the NN SRC has distribution with the width 
being proportional to $k_F$ \cite{Ciofi_Simula,Jan,Tang}.    Thus one expects the  accuracy of the observed relations 
((\ref{p=n}) and (\ref{highn}))  to be worsen with increase of A.

However, due to  the mean field character of the C.M. motion as well as the equal 
contributions of  the $pp$ and $nn$ SRCs to the overall strength of the NN correlations one expects the validity of modified relation:
 \begin{equation}
x_p^\gamma \cdot n^{A}_{p}({p}) \approx x_n^\gamma\cdot n^A_n({p}),
\label{p=n_cm}
\end{equation}
where  $\gamma\equiv \gamma(k_F) \lesssim 1$, with $\gamma$ decreasing with an increase of $A$ (or $k_F$). 
The same $\gamma$ factor will enter also in the high momentum 
part of the momentum distribution of the protons and neutrons
\begin{equation}
 n^{A}_{p/n}({p}) \approx {1\over (2 x_{p/n})^\gamma} a_2(A,y)\cdot n_d({p}),
 \label{highn_cm}
\end{equation}
which will diminish the imbalance between high momentum protons and neutrons presented  in Table~1. 

Very recently, the above predictions have been checked for momentum distributions of 
asymmetric infinite nuclear matter at above saturation densities calculated within Green function method\cite{Rios,WimD}.  
These calculations  observe the scaling of the weighted ratios of the high momentum parts of the proton and neutron momentum 
distributions and indicate that  the  power law  scaling behavior of  Eq.(\ref{p=n_cm})  valid for moderate asymmetries.
 This and the above discussed experimental measurements of  
 $^{56}Fe$ and $^{208}Pb$  are  the first indications that predictions of Eqs.(\ref{p=n},\ref{highn}) or (\ref{p=n_cm},
\ref{highn_cm}) may have  validity for  heavy nuclei and infinite nuclear matter. 

Overall, the realistic nuclear structure calculations  that 
can systematically incorporate short range correlations for asymmetric nuclei(see e.g. \cite{Alvioli,Rios})  combined 
with experimental studies of $A(e,e'n)X$  reactions   will allow  to check the predictions of 
Eqs.(\ref{p=n_cm},\ref{highn_cm}) as well as evaluate the $\gamma$ factor as a function of  nuclear  parameters.

\section{Possible Implications and  universality of the predicted features for two-compnent Fermi systems}
The implications of the above made observations could range from the EMC effects to the proton properties 
in high density asymmetric nuclear matter. These observations suggest several   new directions  in studies of the high momentum component of 
asymmetric nuclei.

-For example; combining three following observations: (i) nuclear  medium modification (EMC effect)
of parton distribution functions~(PDFs)   are proportional to the virtuality (momentum) of the bound 
nucleon (see e.g. \cite{FS88,tagged,nestfun,hnm}); 
(ii) high momentum protons dominate in neutron rich  nuclei (this article) and; 
(iii) PDFs of  proton dominate that of the neutron at $x_{Bjorken}\ge 0.3$ (see e.g. \cite{PDG}), 
one can conclude that the EMC effects for neutron rich nuclei will be defined  mainly by the proton component in the nucleus.
This may explain\cite{memc} the large A part of the recently observed correlation between the strengths of the EMC  and SRC 
effects\cite{Larry1,Larry2}.   

- The prediction of  the enhanced contribution of protons in the EMC effect  indicates that in average 
the $u$-quarks will be more modified than $d$-quarks  in 
neutron rich nuclei and the effect will  grow with  A. This 
provides  an alternative explanation\cite{aps} of the NuTeV anomaly\cite{Zeller,Cloet}.  The predicted effect also 
can be  checked in parity violating deep inelastic scattering off the heavy nuclei.

- The discussed new features of  the high momentum component of nucleon momentum distributions  
could be relevant also for high density asymmetric nuclear matter.  In Ref.\cite{proa2} such a possibility is discussed for 
neutron stars at the cooling threshold of  direct neutrino scattering
(referred to as URCA processes) with  $x_{p}\sim {1\over 8}$ and $y \sim {7\over 9}$.  For example 
it is observed\cite{proa2,pnm} that if the above made observations are valid for infinite nuclear matter then 
starting at  three nuclear saturation densities,  protons will predominantly 
populate the high momentum part of the momentum distribution.  This may have an implication for several properties 
of neutron stars such as cooling through the direct  URCA processes, superfluidity of protons,  the magnetic field of the 
stars as well as the distribution of protons in   the core  of the  massive neutron stars.

\medskip
\medskip

Our observations  in this work follow  from two main general conditions: First, that the interaction is short range and in high 
momentum limit the multiparticle wave function can be factorized to NN correlated and A-2 mean field components. 
Second: the $pn$ interaction significantly dominates that of the  $pp$ and $nn$ interactions.  

As such, the present  results may have  a relevance to any 
asymmetric two-component Fermi system for which above two conditions are satisfied:
that is  the interaction within  each component is suppressed 
while the mutual interaction between two components is finite and short range.  
In such a situation, according to our observations the momentum distribution of the small component will 
be shifted to the high momentum part of  the distribution.

 It is interesting that the similar situation potentially can be 
realized in  two-fermi-component ultracold atomic systems\cite{Shin:2006zz}  but with the mutual s-state 
interaction.
One of the most intriguing aspects of   such systems is that in the  large asymmetric limit  they exhibit very rich 
phase structure with indication of the  strong modification of the small component of the mixture\cite{Bulgac1,Bulgac2}. In this 
respect our  case may be  similar to that of ultracold atomic systems, with the difference that the 
interaction between components has a tensor nature.\\

\section{Summary and Conclusions:}
Based on  the dominance of the tensor forces in the NN system for the momentum range of $\sim k_F-600$MeV/c
we observe  a new scaling relation between $p$- and $n$-
 high momentum distributions weighted by 
their fractions in the nuclei (Eq.\ref{p=n}). Using this,  together with  their relation to the high momentum 
distribution of the deuteron  we  arrive at the  second observation,  according to which the strengths of 
the $p$- and $n$- high momentum components  are inversly proportional to their relative fractions.
Based on these observations  we constructed the $p$- and $n$-
high momentum distributions  for asymmetric nuclei and estimated the overall fraction of nucleons being 
in the high momentum part of the momentum distribution.

The validity of our observations for light nuclei are  confirmed by direct calculations using realistic wave functions.
The first experimental measurements  for large A nuclei and calculations for infinite nuclear matter  indicate the relevance 
of the predictions also for heavy nuclei and nuclear matter. 

We  also observe  that the effects due to center of mass motion of NN SRCs as well as contributions from $pp$ and $nn$ 
SRCs  will diminish the estimated imbalance between high momentum protons and neutrons for large  large A nuclei.  
If this imbalance will be observed for heavy  nuclei and infinite nuclear matter it will have  multitude of implications,  
some  of  which we discussed in the text.

The author is  thankful to Drs.~J.~Arrington,  W.~Boeglin, A. Bulgac, W.~Dickhoff, L.~Frankfurt, G.~Garvey, O.~Hen, D.~Higinbotham, S.~Pastore, E.~Piasetzky,  A.~Rios, M.~Strikman and  L.~Weinstein  for helpful comments and discussions. Special thanks to 
Drs.~R.~Wiringa, T.~Neff  and  W.~Horiuchi for providing the results of their calculations.
This work is supported by U.S. DOE  grant under contract DE-FG02-01ER41172.

\end{document}